\documentstyle[epsfig]{elsart}
\begin{document}
\begin{frontmatter}

\hfill
$\mbox{\small{\begin{tabular}{r}
${\rm Napoli-DSF-99-10}$
\end{tabular}}}$

\vspace{1cm}

\title{The semileptonic form factors of B and D mesons in the
Quark Confinement Model}
\author[bltp]{M.A. Ivanov,}
\author[infn]{P. Santorelli,}
\author[infn]{N. Tancredi}
\address[bltp]{Bogoliubov Laboratory of Theoretical Physics, \\
Joint Institute for Nuclear Research, 141980 Dubna, Russia}
\address[infn]{Dipartimento di Scienze Fisiche, \\
Universit\`a "Federico II" di Napoli, Napoli, Italy\\
and\\
INFN Sezione di Napoli}
\begin{abstract}
The form factors of the weak currents, which appear in the semileptonic
decays of the heavy pseudoscalar mesons are calculated within the quark
confinement model by taking into account, for the first time, the
structure of heavy meson vertex and the finite quark mass contribution
in the heavy quark propagators. The results are in quite good agreement
with the experimental data.
\end{abstract}
\end{frontmatter}

\noindent
\section{Introduction}

The study of semileptonic decays of heavy pseudoscalar mesons can be
used to determine the elements of the Cabibbo-Kobayashi-Maskawa (CKM)
matrix. The decay $D\to K(K^*) l\nu$ is related to $|V_{cs}|$, $B\to
D(D^*)l\nu$ and $B\to \pi(\rho)l\nu$ are proportional to $|V_{cb}|^2$
and $|V_{ub}|^2$, respectively. In the charm sector, however, the CKM
elements can be determined independently of the D semileptonic decay
rate using unitarity of the CKM matrix and the smallness of $V_{cb}$
and $V_{ub}$ \cite{PDG,Rich,Stone}. Thus, the theoretical
predictions for the form factors and their $q^2$-dependence
can be tested.

The study of heavy-to-heavy transitions in decays of $B\to D(D^*)l\nu$
is considerably simplified by the spin-flavor symmetry \cite{IW}. In
the limit of infinite quark mass, in fact, the quark mass and spin
decouple from the dynamics of the decay, leading to numerous symmetry
relations among form factors which can all be related to a single universal
form factor, the Isgur-Wise function. At zero recoil, this function is
known to be normalized to unity, which allows one to determine
$|V_{cb}|$ from the measured $B\to D^*l\nu$ spectrum in the small region
near the zero recoil point. The theoretical symmetry corrections are of
$1/m_Q^2$ order due to the Luke's theorem \cite{Luke}.

The determination of $|V_{ub}|$ from analysis of $B\to\pi(\rho) l\nu$
decays is one of the most important and challenging measurements in
B-physics since the rate for these decays is expected to be only about
$1\%$ of the inclusive semileptonic decay rate. The exclusive
calculations for $B\to X_u l\nu$ are more difficult than those for $B\to
X_c l\nu$, because the range of recoil velocities available to the light
final-state mesons is much larger than for the charm mesons. One
therefore expects a much larger variation in the form factors, which are
still poor known, that enter into the decay rate. As a result,
measurements of $|V_{ub}|$ are currently quite model dependent, and
there is substantial variation among values obtained using different
models \cite{Rich,Stone}.

The main goal of the present paper is to describe the heavy-to-heavy and
heavy-to-light transitions within the quark confinement model (QCM)
\cite{QCM}, by taking into account for the first time the nonlocal
heavy-light quark vertices.

The QCM approach is based on modelling the confined light quarks with
the assumption of {\it local} hadron-quark coupling. It successfully
describes many static and non-static properties of light hadrons. The
extension of this approach to heavy quark physics has been done in
\cite{IKM} by assuming that the free Dirac propagators can be employed
for charm and bottom quarks. It might be justified by the observation
that heavy quarks weakly interact with vacuum background fields, and
therefore they can be considered as free particles with large
constituent masses. The scaling laws for leptonic decay constants and
semileptonic form factors are reproduced in the heavy quark limit. In
addition, the Isgur-Wise function has been calculated. However, the
Isgur-Wise function is larger than in other approaches and in the fitted
experimental data. In \cite{IM,IV} the infrared behavior of the heavy
quark has been taken into account by modifying its conventional
propagator in terms of a single parameter $\nu$ and the heavy-to-light
form factors have been calculated.  In this paper we introduce the
vertex function describing the distribution of constituents inside a
heavy meson. Such distribution is related to the heavy-meson
Bethe-Salpeter amplitude in the approach based on the Dyson-Schwinger
equations \cite{DSE}

\section{Model}
\baselineskip 20pt

The QCM approach \cite{QCM} is based on the effective interaction
Lagrangian for the transition of hadron into quarks:
\begin{equation}
\label{lag}
{\cal L}_{{\rm int}} (x)=g_H H(x) \int\!\! dx_1 \!\!\int\!\! dx_2
\Phi_H (x;x_1,x_2)
\bar q(x_1) \Gamma_H \lambda_H q(x_2)\,.
\end{equation}
Here, $\lambda_H$ and $\Gamma_H$ are the Gell-Mann and Dirac matrices,
respectively, which provide the flavor and spin numbers of mesons $H$.
The function $\Phi_H$ is related to the scalar part of Bethe-Salpeter
amplitude. The local form $\Phi_H(x;x_1,x_2)=\delta(x-(x_1+x_2)/2)
\delta(x_1-x_2)$ has been used in the QCM \cite{QCM}.

The coupling constants $g_H$ defined by what is usually called the
{\it compositeness condition} proposed in \cite{SW} and extensively
used in \cite{QCM}, is given by
\begin{equation}
\label{z_h}
Z_H=1-\frac{3g^2_H}{4\pi^2}\tilde\Pi^\prime_H(m^2_H)=0
\end{equation}
where $\tilde\Pi^\prime_H$ is the derivative of the meson mass operator.

In the QCM-approach the light quark propagators are given
by an entire (non-pole) function to ensure the quark confinement:
\begin{equation}
\label{averaging}
{1\over m_q-\not\!p}\,\,\Rightarrow \,\,
\int\!\!{d\sigma_\mu \over\Lambda \mu-\not\! p}=G(\not\! p)=
{1\over\Lambda}\biggl[a(-{p^2\over\Lambda^2})+
{\not\! p\over\Lambda} b(-{p^2\over\Lambda^2})\biggr]
\end{equation}
with the functions $a$ and $b$ defined by

\begin{equation}
\label{fk}
a(-z)=\int\!\!{\mu d\sigma_\mu \over \mu^2-z}
\hspace{1.5cm}
b(-z)=\int\!\!{d\sigma_\mu \over \mu^2-z}.
\end{equation}

Moreover, to conserve the local properties of Feynman diagrams like the
Ward identities, one has the following prescription for the
modification of a line with n-light quarks within the Feynman diagram
\cite{IV}:
\begin{equation}
\label{pod}
\prod\limits_{i=0}^n \frac{1}{m_q-\not\! p_i}\Gamma_i
\Rightarrow
\int\!\! d\sigma_\mu \prod\limits_{i=0}^n
\frac{1}{\Lambda \mu-\not\! p_i}\Gamma_i\,.
\end{equation}
It is useful to introduce the notation

\begin{eqnarray}
&&\sigma(\not\! k)\equiv \sigma_S(-k^2)+\not\! k \sigma_V(-k^2)
\hspace{1cm}
\sigma_S(z)\equiv \frac{\mu}{\mu^2+z} \hspace{1 cm}
\sigma_V(z)\equiv \frac{1}{\mu^2+z}
\nonumber\\
&&\nonumber\\
&&\int d\sigma_\mu \sigma_S(z)=a(z) \hspace{1cm}
\int d\sigma_\mu \sigma_V(z)=b(z)
\nonumber\\
&&\int d\sigma_\mu \sigma_S(z_1)\sigma_V(z_2)=
\int d\sigma_\mu \sigma_V(z_1)\sigma_S(z_2)=
-\frac{a(z_1)-a(z_2)}{z_1-z_2}
\nonumber\\
&&\int d\sigma_\mu \sigma_V(z_1)\sigma_V(z_2)=
-\frac{b(z_1)-b(z_2)}{z_1-z_2}
\nonumber\\
&&\int d\sigma_\mu \sigma_S(z_1)\sigma_S(z_2)=
\frac{z_1b(z_1)-z_2b(z_2)}{z_1-z_2}
\nonumber\\
&&
\int d\sigma_\mu \sigma_V(z_1)\sigma_V^\prime(z_2)=
-\frac{[b(z_1)-b(z_2)]-(z_1-z_2)b^\prime(z_2)}{z_1-z_2}\,.
\nonumber
\end{eqnarray}
where the confinement functions employed in \cite{QCM} have the forms:
\begin{equation}
a(u)=a_0\exp(-u^2-a_1u) \hspace{1cm} b(u)=b_0\exp(-u^2+b_1u)\, .
\end{equation}
The following values for the free parameters $a_i$,
$b_i$, and $\Lambda$:
$$
a_0=b_0=2 \hspace{1cm} a_1=1 \hspace{1cm} b_1=0.4
\hspace{2cm}  \Lambda=460 \; {\rm MeV};
$$
give a good description of the hadronic properties at low
energies \cite{QCM}.

The hadron-quark coupling constants for light, pseudoscalar and vector,
mesons and heavy pseudoscalar mesons are also determined from the
compositeness condition \cite{QCM} and written down
\begin{equation}\label{norm_L}
g_P={2\pi \over \sqrt 3}\sqrt{2 \over R_P(m_P)},\quad\quad
R_P(x)=B_0+{x\over 4}\int\limits_0^1\!\! du b(-{ux\over 4})
{(1-u/2)\over\sqrt{1-u}}\,.
\end{equation}
Note that from now on all masses and momenta in the structural
integrals are given in units of $\Lambda$.

The heavy quark propagator is given by

\begin{equation}\label{heavy}
S_Q(k+p)=\frac{1}{M_Q-\not\! k- \not\! p}\,.
\end{equation}

\section{Form factors}

We consider the leptonic  $H(p)\to l\nu$, semileptonic heavy-to-heavy
$B(p)\to D(p^\prime) l\nu$ and semileptonic heavy-to-light
$H(p)\to P(p^\prime) l\nu$ decays, where $H(p)$ represents a $B$ (or
$D$) meson with momentum $p$ ($p^2=m^2_H$) and $P(p^\prime)$ can be
a $\pi$ or $K$ meson with momentum $p^\prime$ ($p^{\prime 2}=m^2_P$).
The invariant amplitudes describing the decays are:
\begin{eqnarray}
A(H(p)\to e \nu)&=&
{ G_F \over \sqrt{2} }
V_{Qq}
(\bar e O_{\mu}\nu) M_H^\mu(p)\\
A(B(p)\to D(p') e\nu)&=&{G_F\over \sqrt{2}}V_{bc}(\bar e O_{\mu}\nu)
M^{\mu}_{BD}(p,p^\prime)\\
A(H(p)\to P(p')e\nu)&=&{G_F\over \sqrt{2}}V_{Qq}(\bar e O_{\mu}\nu)
M^\mu_{HP}(p,p^\prime),
\end{eqnarray}
where $G_F$ is the Fermi weak-decay constant, $V_{Qq}$ is the appropriate
element of the Cabibbo-Kobayashi-Maskawa matrix ($q$ denotes a light quark
and $Q$ a heavy quark) and the matrix elements of the hadronic currents
are:
\begin{eqnarray}
M_H^\mu(p)&=&
{3\over 4\pi^2}g_H\Lambda^2\!\!\int\!\!{d^4k\over 4\pi^2i}\phi_H(-k^2)
\nonumber\\
&&\times
{\rm tr}\biggl[O^\mu S_Q(\not\! k+\not\! p)\gamma^5 G(\not\! k)\biggr]=
f_H p^\mu\ ,\\
\nonumber\\
&&\nonumber\\
M^{\mu}_{BD}(p,p^\prime)&=&
{3\over 4\pi^2}g_Bg_D\Lambda\!\!\int\!\!{d^4k\over 4\pi^2i}
\phi_B(-k^2)\phi_D(-k^2)
\nonumber\\
&&\times
{\rm tr}\biggl[ S_c(\not\! k+\not\! p^\prime)O^\mu
S_b(\not\! k+\not\! p)\gamma^5 G(\not\! k)\gamma^5 \biggr]
\nonumber\\
&&\nonumber\\
&&=f_+^{BD}(q^2)(p+p^\prime)^\mu + f_-^{BD}(q^2)(p-p^\prime)^\mu\ ,\\
\nonumber\\
&&\nonumber\\
M^{\mu}_{HP}(p,p^\prime)&=&
{3\over 4\pi^2}g_Hg_P\Lambda \!\!\int\!\! d\sigma_\mu
\!\!\int\!\!{d^4k\over 4\pi^2i}\phi_H(-k^2)
\nonumber\\
&&\times
{\rm tr}\biggl[O^\mu S_Q(\not\! k+\not\! p)\gamma^5
\sigma(\not\! k) \gamma^5 \sigma(\not\! k +\not\! p)\biggr]
\nonumber\\
&&\nonumber\\
&&=f_+^{HP}(q^2)(p+p^\prime)^\mu + f_-^{HP}(q^2)(p-p^\prime)^\mu\ .\\
\nonumber
\end{eqnarray}
From the compositeness condition (in Eq. (\ref{z_h})), the expression
for the propagators, in Eqs. (\ref{averaging}), (\ref{fk}) and
(\ref{heavy}), and the method outlined in \cite{DSE}, we obtain for the
heavy decay constants and heavy to heavy form factors 
\begin{eqnarray}
g_H&=&\sqrt{\frac{4\pi^2}{3\,J_3^{(+)}(m_H,m_H)}}
\hspace{1cm}
f_H=\frac{3}{4\pi^2}\;g_H\;J_2(m_H)
\nonumber\\
&&\nonumber\\
f_\pm^{BD}&=&\frac{3}{4\pi^2}\;g_B g_D\;J_3^{(\pm)}(m_B,m_D)
\end{eqnarray}
with
\begin{eqnarray}
J_2(m_H)&=&\int_0^\infty du \frac{u}{(1+u)^2} z^{\prime}\phi_H(z)
\left [\left(1+\frac{u}{2}\right ) a(z)+ \frac{u}{2} M_Q b(z)\right ]
\nonumber\\
J_3^{(+)}(m_H,m_H)&=&\int_0^{\infty}du \frac{u}{(1+u)^3}\phi^2_H(z)
\nonumber\\
&&\times\left [ M_Q a(z)+
\frac{1}{2}b(z)\left(2z+u(m_H^2+M_Q^2+z)\right)\right]
\nonumber\\
&&\nonumber\\
J_3^{(+)}(m_B,m_D)&= &\frac{1}{2}\int_0^1 dx
\int_0^\infty du\frac{u}{(1+u)^3}\phi_B(z_x)\phi_D(z_x)
\left\{
a(z_x)\left(M_b+M_c\right)
\right.
\nonumber\\
&&
\left.
+b(z_x)\left[u\left(M_b M_c+m_D^2(1-x)+x m_B^2+z_x\right)+2z_x\right]
\right\}
\nonumber\\
&&\nonumber\\
J_3^{(-)}(m_B,m_D)& = &\frac{1}{2}\int_0^1 dx
\int_0^{\infty}du\frac{u}{(1+u)^3}\phi_B(z_x)\phi_D(z_x)
\nonumber\\
&&
\left\{
a(z_x)\left[M_c-M_b+2u\left(M_c-x(M_b+M_c)\right)\right]
\right.
\nonumber \\
&&
\left.
+ b(z_x)u\left[(1-2x)(z_x-M_b M_c)+m_D^2(1-x)-xm_B^2\right]
\right\}
\nonumber
\end{eqnarray}
where the variables are given by
\begin{eqnarray}
&&z=uM_Q^2-\frac{ u}{1+u}\;m_H^2
\hspace{1.5cm}
z^{\prime}=M_Q^2-\frac{1}{(1+u)^2}\;m_H^2
\nonumber\\
&&z_x=u\left\{x\left[M_b^2-\frac{m_B^2}{1+u}\right]
         +(1-x)\left[M_c^2-\frac{m_D^2}{1+u}\right]
         -\frac{u}{1+u}x(1-x)q^2\right\}\ .
\nonumber
\end{eqnarray}
For the heavy to light form factors, instead, the analytical expressions
are
\begin{eqnarray}
f_{\pm}^{HP}(q^2)&=&g_H g_P \left[\frac{3}{4\pi^2}\right]\frac{2}{\pi}
\int_0^{\infty}r dr \int_0^{\infty} \frac{d{\alpha}}{(1+\alpha)^3}
\int_{-1}^1 \frac{d{\gamma}} {\sqrt{1-\gamma^2}} \phi_H(z_1)
\ \frac{1}{2}[G_1 \pm G_2]
\nonumber
\end{eqnarray}
where the functions $G_1(z_1,z_2)$ and $G_2(z_1,z_2)$ can be written as
\begin{eqnarray}
G_1&=&F_{SS}(z_1,z_2)+z_1 F_{VV}(z_1,z_2)-2m_P^2t^2F_{VV'}(z_1,z_2)
\nonumber\\
&&\nonumber\\
G_2&=&M_Q(1+u) F_{SV}(z_1,z_2)+\left(z_1(1+u)+u m_H^2 \right)F_{VV}(z_1,z_2)
\nonumber\\
&&+2t^2\left(F_{SS'}(z_1,z_2)+z_1 F_{VV'}(z_1,z_2)\right)
\nonumber
\end{eqnarray}
and
%
%
\begin{eqnarray}
z_1&=&r^2+u M_Q^2-\frac{u m_H^2}{1+u}
\hspace{2cm}
t=r \sqrt{1-\gamma^2}
\nonumber\\
&&\nonumber\\
z_2 &=& x_2 + i y_2 = \left [ r^2+u M_Q^2-\frac{u q^2}{1+u}-\frac{m_P^2}{1+u}
                   \right] + i \left  [ \frac{2r\gamma m_P}{\sqrt{1+u}}
                               \right ]\, .\nonumber
\end{eqnarray}
The functions $F_{II}$ appearing in $G_1$ and $G_2$ are defined as:
$$
F_{SS}(z_1,z_2)\equiv
\int d\sigma_\mu \sigma_S(z_1)\sigma_S(z_2),
\hspace{0.5cm}
F_{VV^\prime}(z_1,z_2)\equiv
\int d\sigma_\mu \sigma_V(z_1)\sigma_V^\prime(z_2),
\hspace{0.3cm}
{\rm etc.}
$$
Before closing this section, we discuss the
behaviour of the heavy-to-heavy form factors in the limit of $M_b,~M_c\to
\infty$. We shall show that our model reproduces, in this limit,
all the scaling laws predicted by the Heavy Quark Effective Theory at
leading order.

In particular, in the heavy quark limit [$m_H^2=(M_Q+E)^2$ and
$M_Q\to\infty $] one finds
\begin{eqnarray}
&&
\frac{3g^2_{H}}{4\pi^2}\cdot \frac{1}{2M_Q}\cdot I_{HH}=1
\hspace{1cm}
I_{HH}=\int\limits_0^\infty du
\phi_H^2(\tilde z)\{a(\tilde z)+\sqrt{u} b(\tilde z) \},
\nonumber\\
&&\nonumber\\
&&
f_P= \Lambda\ \sqrt{\frac{2}{M_Q}}\ \frac{\sqrt{3}}{2\pi}
\sqrt{\frac{1}{I_{HH}}}\ 
\int\limits_0^\infty du (\sqrt{u}-E)\phi_H(\tilde z)
\{a(\tilde z)+\frac{1}{2}\sqrt{u} b(\tilde z) \},
\nonumber\\
&&\nonumber\\
&&
f_{\pm}=\frac{M_Q\pm M_{Q^\prime}} {2\sqrt{M_QM_{Q^\prime}}}
\cdot\xi(w)
\nonumber
\end{eqnarray}
where the Isgur-Wise function, $\xi(w)$, is given by
\begin{equation}
\xi(w)=
\frac{1}{I_{HH}}\cdot
\int\limits_0^1\frac{d\tau}{W}
\int\limits_0^\infty du \phi_H^2(\tilde z_W)
\biggl[a(\tilde z_W)+\sqrt{u/W} b(\tilde z_W)\biggr]
\end{equation}
with
$$
W=1+2\tau(1-\tau)(w-1) \hspace{1cm}
\tilde z_W=u-2E\sqrt{u/W} \hspace{1cm}
\tilde z=u-2E\sqrt{u}\, .
$$
It is readily seen that the upper bound for the Isgur-Wise function
is obtained for $E=0$, namely
\begin{equation}
\xi(w)\le \bar\xi(w)=\xi(w)|_{E=0}
=\frac{1}{1+R}\biggl\{\frac{\ln[w+\sqrt{w^2-1}]}
{\sqrt{w^2-1}}+\frac{2R}{1+w} \biggr\}
\label{ximax}
\end{equation}
where
$$
R=\frac{\int\limits_0^\infty du\ \phi_H^2(u)\ \sqrt{u}\ b(u)}
       {\int\limits_0^\infty du\ \phi_H^2(u)\ a(u)}\, .
$$
As a consequence of Eq.~(\ref{ximax}) the slope parameter has the
lower bound
$$
\rho^2=-\xi^\prime(1)=\frac{1}{3}\biggl[1+\frac{1}{2}\frac{R}{1+R}\biggr]
\ge \frac{1}{3}\,.
$$

In the heavy quark limit ($p^2=(M_Q+E)^2$, $(p^\prime)^2=0$ and
$M_Q\to\infty$) one finds for the heavy-to-light form factors that
\begin{equation}
f_\pm(q^2) \to \frac{g_\pi}{4\pi}\cdot \sqrt{\frac{6}{I_{HH}}}
\int\limits_0^\infty du (\sqrt{u}-E) \phi_H(\tilde z_1)
\int\limits_0^1d\tau \sqrt{M_Q}
\biggl[\frac{1}{M_Q} \tilde G_1\pm \tilde G_2\biggr].
\end{equation}
Here
\begin{eqnarray}
\tilde G_1&=&F_{SS}(\tilde z_1,\tilde z_2)+
\tilde z_1 F_{VV}(\tilde z_1,\tilde z_2)
\nonumber\\
&&\nonumber  \\
\tilde G_2&=&F_{SV}(\tilde z_1,\tilde z_2)+
\tau\sqrt{u}F_{VV}(\tilde z_1,\tilde z_2)
\nonumber
\end{eqnarray}
with the $F_{II}$'s defined before, $ \tilde z_1=u-2E\sqrt{u}$,\
$\tilde z_2=\tilde z_1+
2 X\tau\sqrt{u}$, and
$$
X=vp^\prime=\frac{M_Q}{2}\biggr[1-\frac{q^2}{M_Q^2}\biggl]\, .
$$

At the end point $q^2=q^2_{\rm max}$ ($X=0$) one can reproduce
the well-known relations among form factors in the heavy quark limit

\begin{eqnarray}
&&\frac{ (f_+ + f_-)^B } { (f_+ + f_-)^D} =\sqrt{\frac{m_D}{m_B}}
\hspace{3cm}
\frac{ (f_+ - f_-)^B } { (f_+ - f_-)^D} =\sqrt{\frac{m_B}{m_D}}\,.
\nonumber
\end{eqnarray}

\section{Results and discussion}

The expressions obtained in the previous section for the form factors
and decay constants are valid for any kind of vertex function $\phi_H(-k^2)$.
Here, we choose a Gaussian form $\phi(-k^2)=\exp\{k^2/\Lambda_H^2\}$ in
Minkowski space. The magnitude  of $\Lambda_H$ characterizes the size of
the BS-amplitude and  is an adjustable parameter in our approach. Thus,
we have  four adjustable parameters: $\Lambda_D$ and $\Lambda_B$ plus
the two heavy quark masses, or binding energies $E_D=m_D-M_c$ and
$E_B=m_B-M_b$. The first two are fixed in such a way the form factors
$f_+^{B\pi}(q^2)$ and $f_+^{DK}(q^2)$ are increasing functions of
$q^2$; we choose $\Lambda_D=0.56$ GeV and $\Lambda_B=0.67$ GeV.
The other parameters are fixed by the least-squares fit to the
observables measured experimentally or taken from a lattice simulation
(see asterisks in  Table~\ref{t1}).

The best fit is achieved for $E_D \approx E_B$, thus we choose to fix
$E_D = E_B$ in such a way we have only two free parameters. The best
values are $E_D=E_B=0.554$ GeV and $V_{cb}=0.043$ which is close to the
world-accepted value \cite{PDG}. The resulting values for the heavy to
light form factors at $q^2=0$ are larger those predicted by
other approaches. It should be stressed that these values are
practically fixed by the assumption that the form factor should be
increasing functions of $q^2$. Moreover, there is a strong correlation
between $f_+^{H\to L}(0)$ and the decay constant $f_H$, {\it i.e.}
smaller values for form factors corresponds to small values for decay
constants. The situation changes if no assumptions are done on the $q^2$
behaviour of the form factors.

We plot the the $q^2$-behaviour of the resulting form factors on Fig.1.
For comparison, the vector dominance, pole model is shown: 
\begin{equation}\label{mon}
f_+^{q\to q'}(q^2)=\frac{f_+^{q\to q'}(0)} {1-q^2/m^2_{V_{qq'}}}
\end{equation}
with $m^2_{V_{qq'}}$ being the mass of the lightest $\bar qq'$-vector meson.
We use
$m_{D^*_s}=2.11$ GeV  for $c\to s$,
$m_{B^*}=5.325$ GeV   for $b\to u$,
$m_{B_c^*}\approx m_{B_c}=6.4$ GeV \cite{CDF-Bc}  for $b\to c$ transitions.
The values of $f_+^{qq'}(0)$ are taken from the Table 1. Also we calculate
the  branching ratios of semileptonic decays  by using widely accepted values
of the CKM matrix  elements \cite{PDG}.

A few comments should be done concerning the comparison of our results
with the results of paper \cite{IS} where the weak decays of pseudoscalar
mesons have been described within the relativistic constituent quark
model with free quark propagators. Since there is no confinement in that
model the binding energies have been found to be relatively small:
$E_D=0.20$ GeV and $E_B=0.22$ GeV. Such values provide the absence
of imaginary parts in the physical amplitudes describing the decays
of the low-lying pseudoscalar mesons. However, the excited states
like vector mesons cannot be considered in a self-consistent manner.
The Quark Confinement Model allows us to give the unified description
of physical observables without quark thresholds in the physical
amplitudes and with a minimum set of parameters: the only parameter
$\Lambda=0.460$ GeV, the size of confinement region, for light quark sector
and four extra parameters ($\Lambda_{B,D}$-the sizes of Bethe-Salpeter
amplitudes, and $E_{B,D}$-the binding energies) for heavy quark sector.
As a result, the accuracy of desciption is less than in \cite{IS}
while, the region of application is considerably wider. 

\begin{table}[t]
\caption{Prediction for leptonic decay constants (in GeV),
form factors and ratios. The "Obs." are extracted from
Refs.~\protect \cite{PDG,CLEO-BD,CLEO-Bpi,Flynn,Wittig,Debbio,MILC} 
$(q^2_M = (m_B-m_D)^2)$.
\label{t1}}
\small{
\begin{tabular}{lll|lll}
       & Obs.                & Calc.  &       & Obs.                & Calc.
\\
\hline
$\ast\ f_D$ & 0.191$^{+19}_{-28}$ & 0.165   & 
$\ast\ f_B$ & 0.172$^{+27}_{-31}$ & 0.135 \\
$\ast\ f_+^{DK}(0)$ & 0.74 $\pm$ 0.03 & 0.77  &
 ${\rm Br}(D\to K l\nu)$  & $(6.8\pm 0.8)\cdot 10^{-2}$ & $8.8\cdot 10^{-2}$\\
$\ast\ |V_{cb}|f_+^{BD}(q^2_M)$ & $(5.09 \pm 0.81)\ 10^{-2}$  & 
$5.1 \ 10^{-2}$  &
 ${\rm Br}(B\to D l\nu)$  & $(2.00\pm 0.25)\cdot 10^{-2}$ & $3.5\cdot 10^{-2}$
\\
$\ast\ f_+^{B\pi}(0)$ & 0.27 $\pm$ 0.11 & 0.55  &
 ${\rm Br}(B\to \pi l\nu)$  & $(1.8\pm 0.6)\cdot 10^{-4}$ & $3.3\cdot 10^{-4}$
\\
\hline
\end{tabular}}
\vspace{1truecm}
\end{table}

\vspace{1cm}

\noindent
{\large\bf Acknowledgments}

We appreciate F. Buccella for many interesting discussions and
critical remarks. M.A.I. gratefully acknowledges the hospitality and
support of the Theory Group at Naples University where this work was
conducted.
This work was supported in part by the Russian Fund for Fundamental
Research, under contract number 99-02-17731-a.

\newpage
%


\begin{figure}[t]
\begin{center}
\begin{tabular}{cc}
\epsfig{file=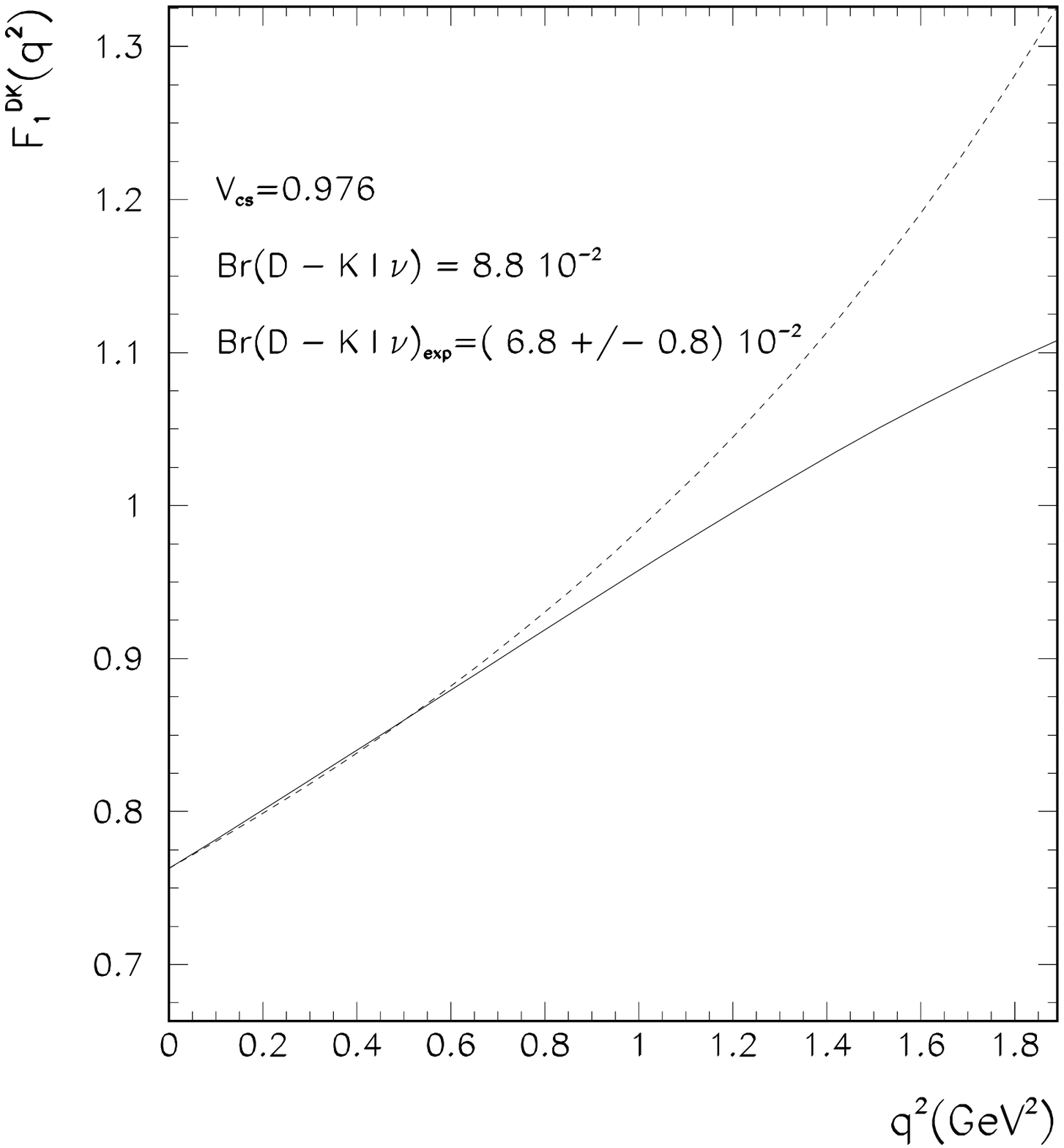,height=8cm} &
\epsfig{file=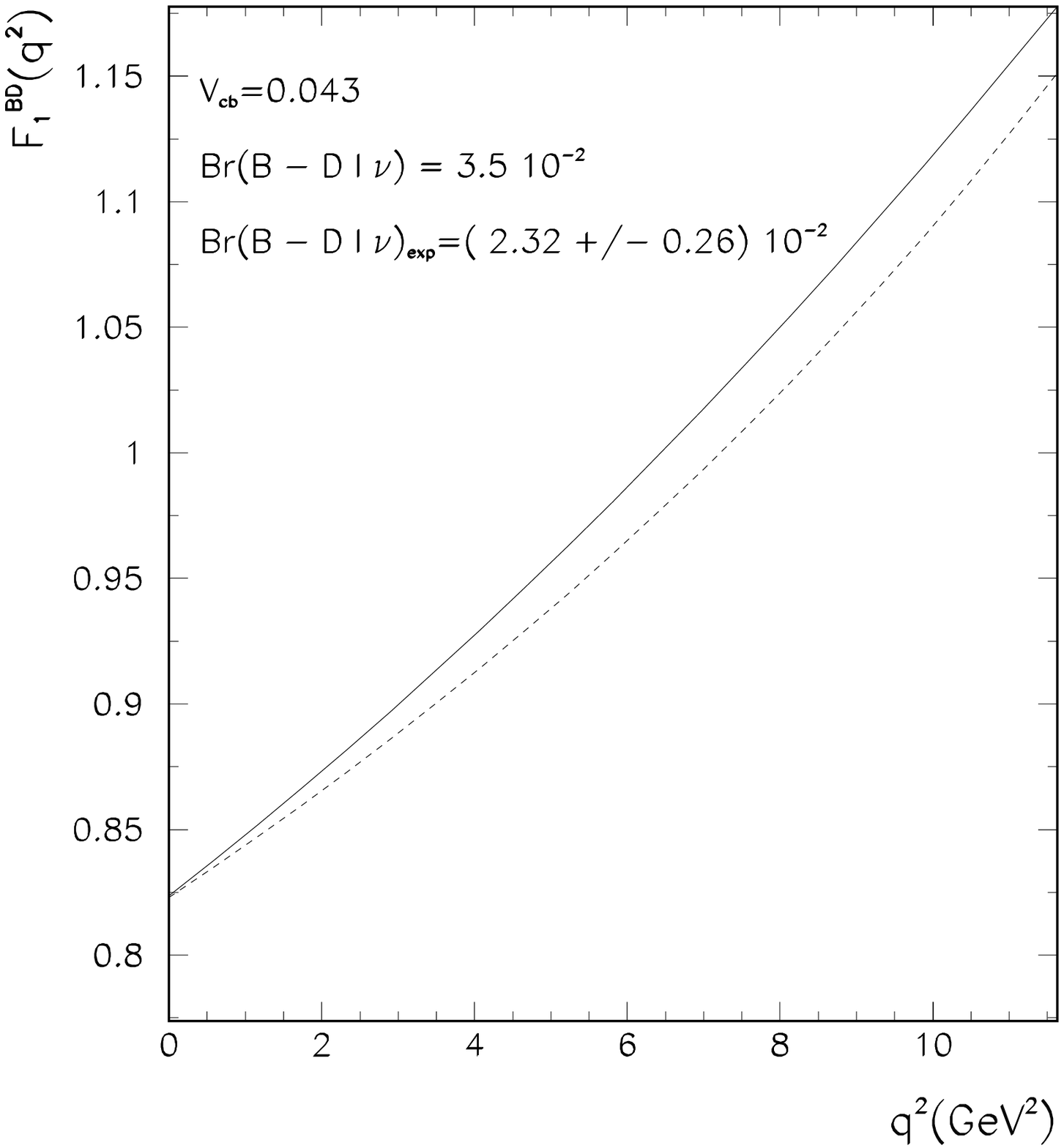,height=8cm} \\
\epsfig{file=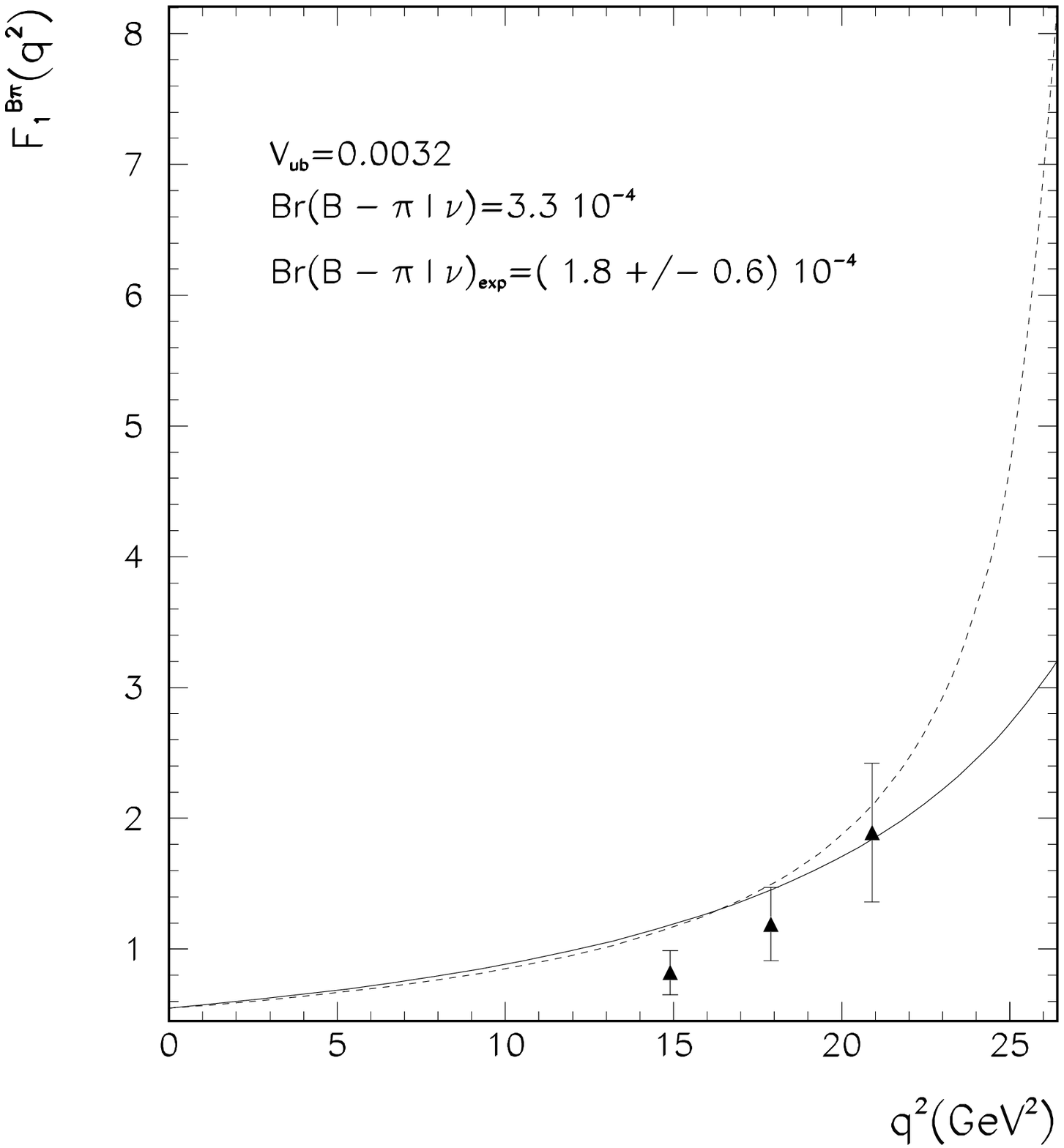,height=8cm} &
\end{tabular}
\end{center}
\caption
{
The semileptonic  $D\to K$, $B\to D$ and $B\to\pi$
form factors with, for comparison, a vector dominance, monopole model
Eq.~(\ref{mon}) and a lattice simulation \protect\cite{Burford}.
Our results: solid lines. Monopole: dotted lines. Lattice: data points.
}
\label{f:fig1}
\end{figure}

\end{document}